\documentclass{article}

\usepackage{PRIMEarxiv}

\usepackage[utf8]{inputenc} % allow utf-8 input
\usepackage[T1]{fontenc}    % use 8-bit T1 fonts
\usepackage{hyperref}       % hyperlinks
\usepackage{url}            % simple URL typesetting
\usepackage{booktabs}       % professional-quality tables
\usepackage{amsfonts}       % blackboard math symbols
\usepackage{nicefrac}       % compact symbols for 1/2, etc.
\usepackage{microtype}      % microtypography
\usepackage{lipsum}
\usepackage{fancyhdr}       % header
\usepackage{graphicx}       % graphics
\graphicspath{{media/}}     % organize your images and other figures under media/ folder

\title{Segmentation-based Information Extraction and Amalgamation in Fundus Images for Glaucoma Detection
}

\author{
  Yanni Wang, Gang Yang\thanks{Corresponding author}  \\
  School of Information \\
  Renmin University of China \\
  Beiing\\
  \texttt{\{yanni.wang, yanggang\}@ruc.educ.cn} \\
   \And
  Dayong Ding, Jianchun Zhao \\
  Vistel AI Lab\\ 
  Visionary Intelligence Ltd.,\\
  Beijing\\
  \texttt{\{dayong.ding, jianchun.zhao\}@vistel.cn} \\
}

\begin{document}
\maketitle

\begin{abstract}
Glaucoma is a severe blinding disease, for which automatic detection methods are urgently needed to alleviate the scarcity of ophthalmologists. Many works have proposed to employ deep learning methods that involve the segmentation of optic disc and cup for glaucoma detection, in which the segmentation process is often considered merely as an upstream sub-task. The relationship between fundus images and segmentation masks in terms of joint decision-making in glaucoma assessment is rarely explored. We propose a novel segmentation-based information extraction and amalgamation method for the task of glaucoma detection, which leverages the robustness of segmentation masks without disregarding the rich information in the original fundus images. Experimental results on both private and public datasets demonstrate that our proposed method outperforms all models that utilize solely either fundus images or masks.

\end{abstract}

% keywords can be removed
\keywords{Glaucoma detection \and Color fundus \and Segmentation mask}

\section{Introduction}
Glaucoma is the leading cause of global irreversible blindness, and the number of people with glaucoma worldwide is projected to reach 111.8 million in 2040 \cite{gw}. However, glaucoma can remain asymptomatic through the early stage, leaving a large number of possibly affected people undiagnosed \cite{ptg}, which if unattended, could lead to a larger population with severe glaucoma and irreversibly damaged vision in the future. Thus, early treatment or intervention is of great importance. However, due to the lack of highly experienced experts for large-scale glaucoma screening, effective automatic glaucoma detection methods are urgently needed.

\begin{figure}[h!]
\centering
\includegraphics[scale=0.4]{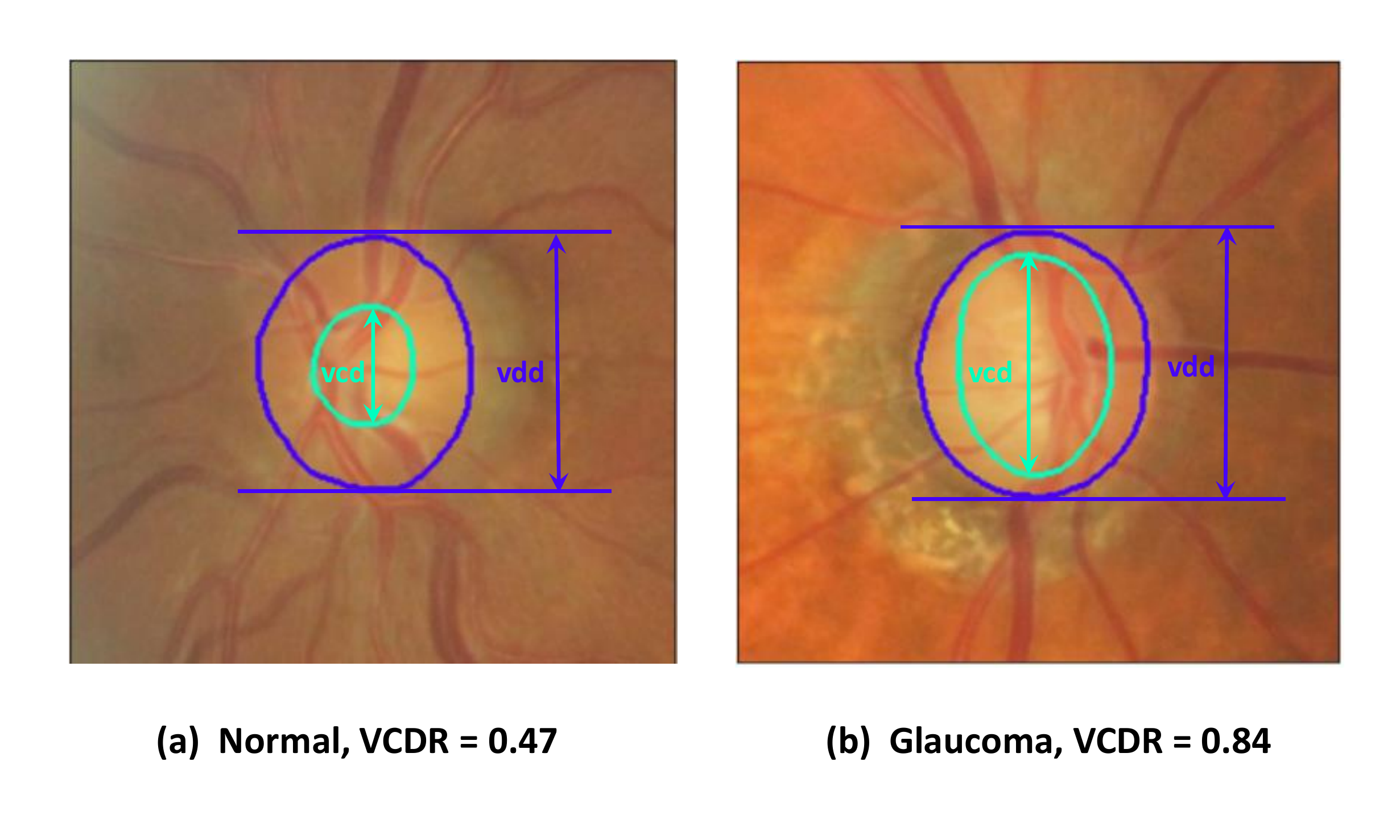}
\caption{%\csentence{Sample figure title.}
  The VCDR is calculated by the ratio of vertical cup diameter(vcd) to vertical disc diameter(vdd). VCDR is highly correlated with glaucoma, and thus accurate segmentation of optic disc and cup becomes a crucial task for glaucoma detection. A larger VCDR generally indicates a higher possibility of glaucoma.}
\label{vcdr}
\end{figure}

\begin{figure*}[t]
  \includegraphics[scale=0.45]{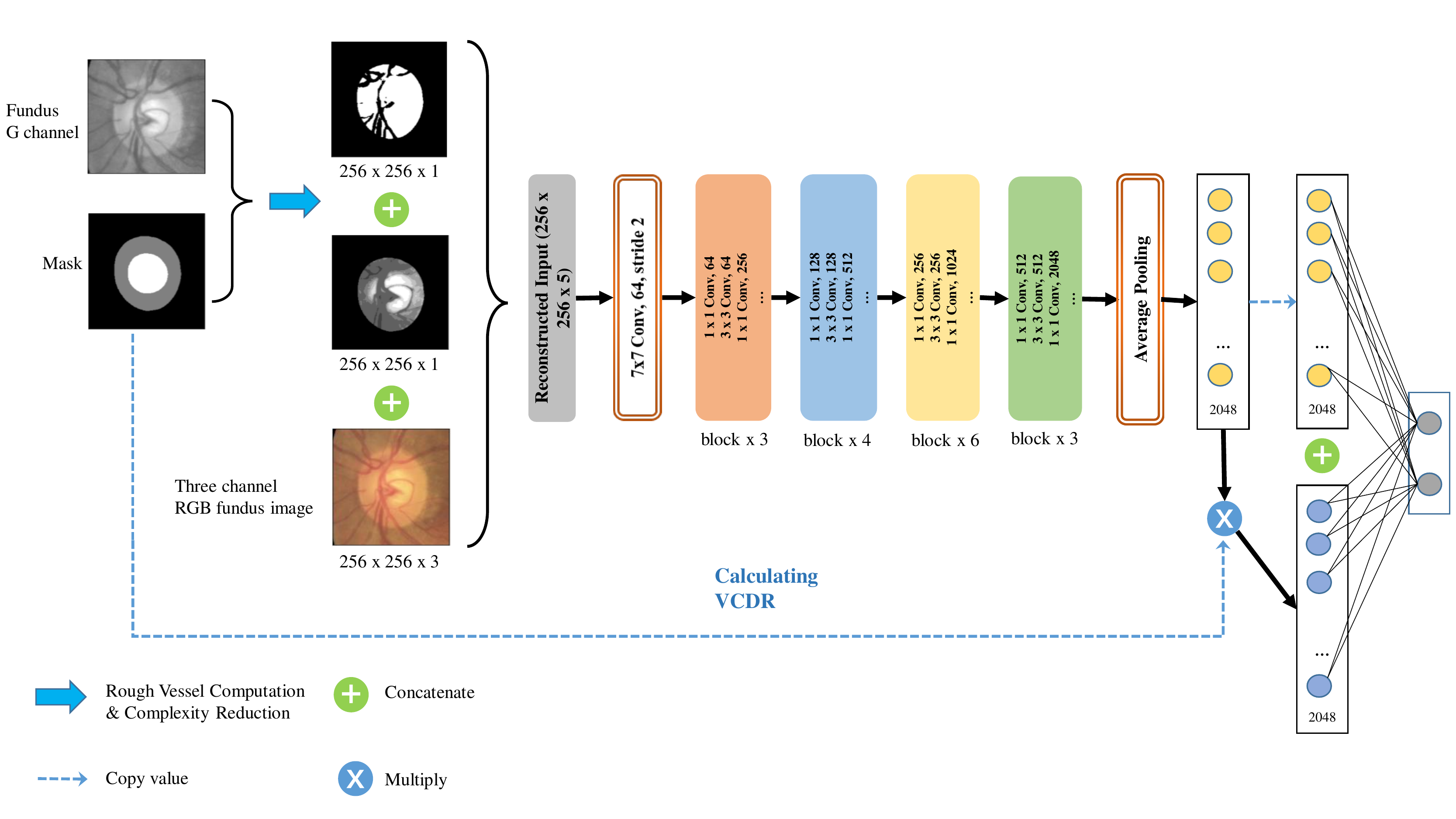}
  \caption{The reconstructed 5-channel input goes through a partially modified Resnet50 structure \cite{resnet} pretrained on ImageNet \cite{ImageNet}. Meanwhile, VCDR values are calculated based on the masks and passed along the network, eventually incorporated into the decision-making pipeline.}
\label{structure}
\end{figure*}

Color fundus photography is a cost effective imaging modality for clinical diagnosis of glaucoma, but it is not easy even for trained ophthalmologists to reliably identify glaucoma in fundus. Moreover, assessment of glaucomatous structural damge in fundus photographs suffers from low reproducibility \cite{m2m}. 

To improve reproducibility and interpretability, and to assist clinicians with diagnosing glaucoma, various clinical measurements are explored \cite{vcdr,bibm1}, among which the vertical cup to disc ratio (VCDR) is well accepted and widely adopted \cite{vcdr}. As a risk factor, a larger VCDR generally indicates a higher possibility of glaucomatous damage \cite{risk}. Due to the fact that VCDR is highly correlated with glaucoma, accurate segmentation of optic disc and cup becomes a crucial task for glaucoma detection, since the segmentation masks are indispensable for the calculation of VCDR, as illustrated in Fig.~\ref{vcdr}.

In recent decades, advances in artificial intelligence algorithms, notably deep learning neural networks, have fueled remarkable progress in medical image analysis \cite{unet,aunet,unet3} and have led to exciting prospects in automating the assessment of glaucoma in color fundus images \cite{review}. Many previous works have proposed to employ deep learning methods for glaucoma detection in fundus photographs, but the segmentation of optic disc and cup is often singled out as an independent task, and the relationship between fundus and segmentation masks in terms of joint decision-making in glaucoma assessment is rarely explored. In most works, the segmentation of optic disc and cup is generally regarded as an standalone upstream task rather than a source of complementary information to fundus images in performing glaucoma classification. Various works have suggested utilizing only the segmentation results to achieve the assessment of glaucoma, regardless of the original information that the color fundus images contain \cite{odc,Mnet}. Though glaucoma classification based solely on disc and cup segmentation results exhibits good generalization ability, there is still considerable room left for improvement, for even VCDRs calculated using ground truth masks only result in baseline level performance \cite{refuge}.

Plenty researches have explored the automatic detection of glaucoma in fundus images using machine learning methods \cite{eff,m2m,da,weak,bibm2}. Li et al. \cite{eff} suggest training a deep learning model based solely on fundus images with labels that multiple graders agree on. Medeiros et al. \cite{m2m} propose a deep learning algorithm trained on fundus images that are labeled with the corresponding global retinal nerve fiber layer thickness measurements from spectral-domain optical coherence tomography (SDOCT). Yu et al. \cite{da} propose to leverage the multi-rater consensus information for the glaucoma classification task. Ghamdi et al. employ semi-supervised transfer learning for glaucoma detection in fundus images. Zhao et al. \cite{weak} propose a weakly-supervised method to achieve evidence identification, optic disc segmentation, and glaucoma diagnosis simultaneously, which only requires weak-label fundus images as training input. 

The segmentation of Optic Disc and Cup is also a well established research field, since vertical cup to disc ratio (VCDR) is considered an indicator of great significance for glaucoma diagnosis \cite{vcdr}. There have been a great number of works which employ machine learning methods to segment optic disc and cup \cite{odc,odcu,semi,bibm3}.

However, in some works, the segmentation of optic disc and cup is singled out as an independent task rather than a source of complementary information for glaucoma detection \cite{weak,semi}. And to another extreme, some works have suggested utilizing only the segmentation results to perform the downstream task of glaucoma assessment, regardless of the rich original information that the color fundus images contain. R. Ali et al. \cite{odc} propose a method for glaucoma screening based on the computed VCDR from the segmentation result, which is obtained through a fuzzy broad learning system. Sevastopolsky proposes a modified U-Net neural network for disc and cup segmentation \cite{odcu}, based on which glaucoma detection is performed.

In this work, we propose to leverage the optic disc and cup segmentation results while exploiting the rich original information in fundus images for glaucoma detection. Not only do we take into consideration the spatial information provided by the segmentation masks, but we also manage to incorporate the VCDR values into the classification pipeline. The main contributions of our work can be described as follows:
\begin{itemize}
\item We propose a simple yet effective method to adaptively compute rough vessel maps, which can serve as a source of complementary information, based on spatial indication provided by the disc and cup segmentation masks.
\item We design an approach to performing complexity reduction by thresholding the values in the fundus images into a small set, but at the same time still retaining the relevant characteristics. 
\item We propose an easy-to-implement method to fuse VCDR values into our deep learning network to improve robustness, fully utilizing relevant ophthalmologic prior knowledge.
\item For the task of glaucoma classification, we manage to leverage the robustness of segmentation-based methods without disregarding the rich information in the original fundus images, resulting in a better performance as well as generalization abilities. Experimental results reveal that our method provides great improvements on both fundus-based and mask-based methods.
\end{itemize}

\section{Methods}
\label{sec:headings}
% \lipsum[4] See Section \ref{sec:headings}.
In this section, we give a detailed description of our proposed method, the overall structure of which is illustrated in Fig.~\ref{structure}. Based on the fundus images and segmentation masks (We employ the HANet method proposed in \cite{hanet} to obtain segmentation results beforehand.), we firstly perform rough vessel map computation and complexity reduction to acquire two new channels of input, which are subsequently concatenated to the original RGB channels of fundus images. The reconstructed 5-channel input then goes through a partially modified Resnet50 structure \cite{resnet} pretrained on ImageNet \cite{ImageNet}. Meanwhile, VCDR values are calculated based on the masks and passed along the network, eventually incorporated into the decision-making pipeline.

\subsection*{Rough Vessel Map Computation}

We propose a simple yet effective method to adaptively compute rough vessel maps, which can serve as a source of complementary information, based on spatial indication provided by the disc and cup segmentation results. 

\begin{figure}[h!]
\centering
\includegraphics[scale=0.55]{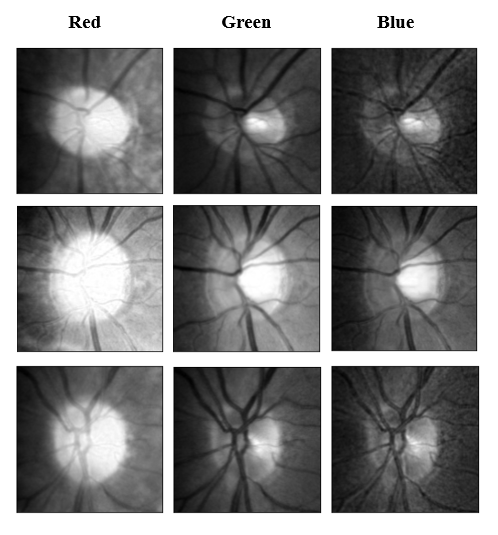}
\caption{The vessel boundaries are visibly more clear and obvious on the green channel than on the red and blue channels of fundus images.}
\label{RGB}
\end{figure}

We choose to utilize only the green channels of color fundus images as the source information to compute rough vessel maps, since the vessel boundaries are visibly more clear and obvious on the green channel than on the red and blue channels of fundus images, as can be observed in Fig.~\ref{RGB}, and thus more suitable for thresholding to obtain the rough vessel maps.

We denote $S_b=\{(x_1,y_1),\ldots\}$ the set of coordinates in the background of the segmentation mask, where it belongs to neither cup nor disc region, and $|S_b|$ is the size of the set. We define a $V$ function that translates the coordinate $(x,y)$ into its pixel value in the green channel of the color fundus image. We calculate the mean pixel value of the background region (suggested by the segmentation results) in the green channel and name it $Back$. Similarly, we define $Rim$ as the mean pixel value of the rim region in the green channel, the set of coordinates in which is denoted as $S_r$, where $|S_r|$ represents the size of the set.

\begin{equation}
Back = \frac{1}{|S_b|} \sum_{(x,y) \in S_b} V(x,y) \label{Back} 
\end{equation}

\begin{equation}
Rim = \frac{1}{|S_r|} \sum_{(x,y) \in S_r} V(x,y) \label{Rim} 
\end{equation}

\begin{figure}[h!]
\centering
\includegraphics[scale=0.55]{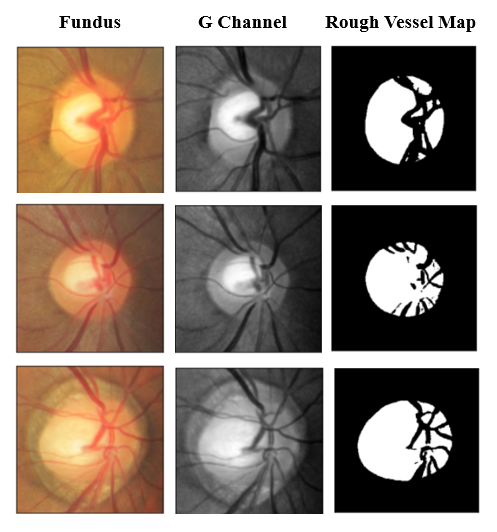}
\caption{We adaptively calculate threshold values to perform global thresholding on the green channels of color fundus images and paint the background region (suggested by the segmentation results) black to reduce noise.}
\label{rv}
\end{figure}

To obtain the rough vessel maps, we adaptively calculate threshold values, denoted as $T_v$ in Eq. (\ref{T_v}), to perform global thresholding on the green channels of color fundus images. To reduce irrelevant noise, in the rough vessel maps, we paint the background region (suggested by the segmentation results) black, as shown in Fig.~\ref{rv}.

\begin{equation}
T_v = Back + \frac{(Rim - Back)}{2} \label{T_v}  
\end{equation}

\begin{figure}[t]
\centering
\includegraphics[scale=0.45]{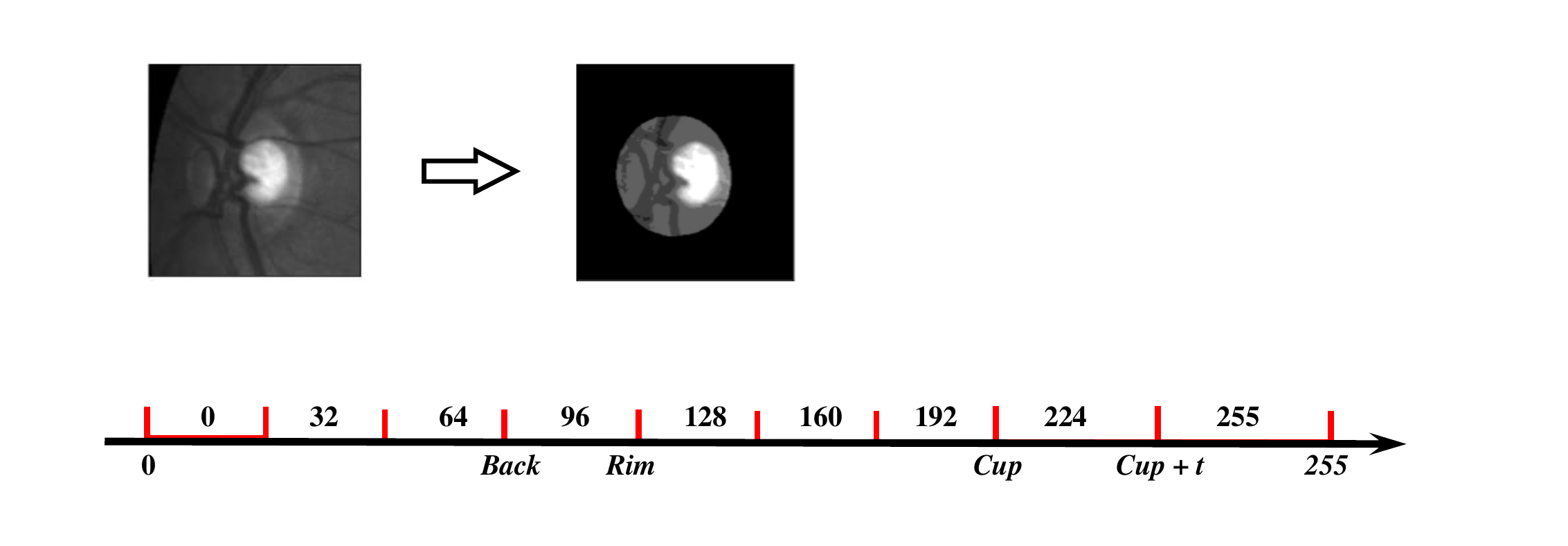}
\caption{After the complexity reduction is perpormed, we paint the background region (suggested by the segmentation results) black to reduce noise.}
\label{cr}
\end{figure}

\subsection*{Complexity Reduction}
We perform complexity reduction by thresholding the green channels of color fundus images into a small set of pixel values, during which the ranges for thresholding are computed adaptively using the indication provided by the segmentation results. On top of $Back$ and $Rim$, we further compute $Cup$ by the same token, which is defined as follows:
\begin{equation}
Cup = \frac{1}{|S_c|} \sum_{(x,y) \in S_c} V(x,y) \label{Cup} 
\end{equation}
where the set of coordinates in the cup area is denoted as $S_c$, and the size of the set $|S_c|$. Moreover, we define a parameter $t$, and we set it to 20 during our experiments. 

As shown in Fig.~\ref{cr}, we set multiple milestone values based on the calculated $Back$, $Rim$, $Cup$, and $t$ for the green channels of fundus images. Then we constrain pixel values within the ranges of adjacent milestones into certain fixed values, which are roughly evenly distributed across 0 to 255. To reduce irrelevant noise, we paint the background region (suggested by the segmentation results) black.

\subsection*{VCDR Incorporation}
To fuse the VCDR into the deep learning network, we construct an additional branch of 2048 neurons by multiplying the existing neuron values before the fully connected layer of a Resnet50 backbone by VCDR. And then we concatenate the reconstructed branch of neurons to the original ones, resulting in a total of 4096 neurons before the fully connected layer, as illustrated in the right corner of Fig.~\ref{structure}.

\subsection*{Datasets}\label{AA}
\begin{itemize}
\item SY   

We construct a private dataset named SY for our experiments. The SY dataset consists of a total of 11470 color fundus images. As shown in Table.~\ref{SY}, there are far fewer glaucoma cases than non-glaucoma cases in SY, which makes it an imbalanced dataset. Meanwhile, although SY is a large dataset, not all images in it are of good quality due to complicated clinical circumstances under which the images were collected. Thus, SY can well represent the problem of long-tail distribution as well as relatively low image quality in the wild.

\item REFUGE

REFUGE is a public dataset released during the Retinal Fundus Glaucoma Challenge \cite{refuge}. It contains a total of 1200 fundus images with ground truth segmentations and clinical glaucoma labels. The ratio of glaucoma cases to non-glaucoma is 1:9, as shown in Table.~\ref{REFUGE}. All images in the training set contains all the images acquired with the Zeiss Visucam 500 camera. In contrast, the validation and test sets include lower resolution images captured with the Canon CR-2 device.
\end{itemize}
\begin{table}[h!]
\caption{Details about the SY Dataset}
\begin{center}
\scalebox{1.2}{
\begin{tabular}{|c|c|c|c|}
\hline
% \textbf{Table}&\multicolumn{3}{|c|}{\textbf{Table Column Head}} \\
\cline{2-4} 
\textbf{Subset} & \textbf{\textit{Glaucoma}}& \textbf{\textit{Normal}}& \textbf{\textit{Total}} \\
\hline
Train&946 & 8222&9168 \\
\hline
Val&119 & 1027& 1146\\
\hline
Test& 119& 1037& 1156\\
\hline
\end{tabular}}
\label{SY}
\end{center}
\end{table}

\begin{table}[h!]
\caption{Details about the REFUGE Dataset \cite{refuge}}
\begin{center}
\scalebox{1.2}{
\begin{tabular}{|c|c|c|c|}
\hline
% \textbf{Table}&\multicolumn{3}{|c|}{\textbf{Table Column Head}} \\
\cline{2-4} 
\textbf{Subset} & \textbf{\textit{Glaucoma}}& \textbf{\textit{Normal}}& \textbf{\textit{Total}} \\
\hline
Train& 40& 360& 400\\
\hline
Val& 40& 360&400 \\
\hline
Test& 40&360 &400 \\
\hline
\end{tabular}}
\label{REFUGE}
\end{center}
\end{table}

\section*{Results and Discussion}
In this section, we demonstrate our experimental details as well as our results on a private dataset and a public dataset.

\subsection*{Evaluation Metrics}

Receiver operating characteristic curve
(ROC) and area under ROC (AUC) are adopted as the primary metrics for evaluating the performance of our glaucoma classification method compared to other baselines.

TP, TN, FP, and FN are the numbers of the true positive glaucoma, true negative glaucoma, false positive glaucoma and false negative glaucoma respectively. Sensitivity and specificity are defined as follows:

\begin{equation}
Sensitivity = \frac{TP}{TP + FN} \label{sen}  
\end{equation}

\begin{equation}
Specificity = \frac{TN}{TN + FP} \label{spe}  
\end{equation}

Since sensitivity and specificity are more relevant from a medical standpoint, we employ the harmonic mean of sensitivity and specificity as our F1-score, defined as follows: 

\begin{equation}
F1 = 2 \times \frac{Sensitivity \times Specificity}{Sensitivity + Specificity} \label{f1}  
\end{equation}

F1-score, sensitivity, and specificity are considered as our secondary metrics.

\subsection*{Implementation Details}
We implement our network with PyTorch on TITAN Xp GPU. Firstly, we crop the disc regions from fundus images and obtain regions of interests (ROIs) of 256$\times$256. We employ the HANet method proposed in \cite{hanet} to obtain segmentation results of these ROIs. The augmentation strategies we adopt include random horizontal flip and random Gaussian blur. Moreover, we employ a Resnet50 \cite{resnet} structure that is pretrained on ImageNet \cite{ImageNet}. We train the SY dataset for 50 epoches and REFUGE for 100, during which the models with the best AUCs are saved.

\subsection*{Baseline Models}
To verify the effectiveness of different modules of our method, we adopt multiple baselines in comparison with our proposed model. Namely, $Fundus_{vcdr}$ stands for the model that utilizes the three-channel RGB fundus images as input and also incorporates our proposed VCDR module. $Fundus$ stands for only using fundus images without involving VCDR. By the same token, $Mask_{vcdr}$ employs three channels of repeated masks as well as the VCDR module, and $Mask$ does not involve VCDR. Moreover, the $VCDR$ models in the last row of Table.~\ref{tab_SY} and Table.~\ref{tab_REFUGE} are logistic regression models trained, validated, and tested solely with the calculated VCDR values. Except for the $VCDR$, all other deep learning models adopt the same Resnet50 backbone.

\begin{figure}[t]
\centering
\includegraphics[scale=0.4]{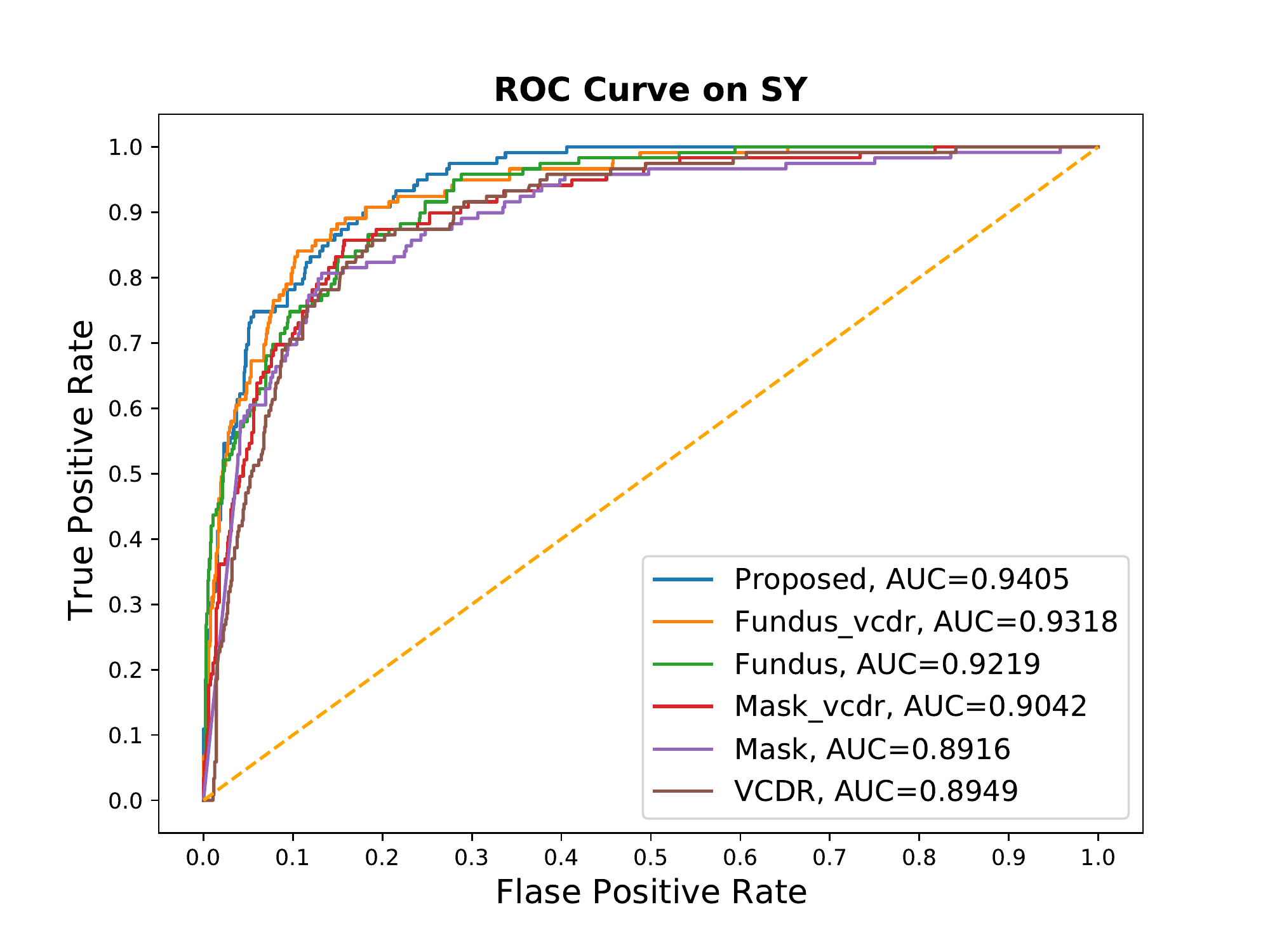}
\caption{ROC curve on SY}
\label{SY_roc}
\end{figure}

\begin{table}[t]
\caption{Experimental Results on SY}
\begin{center}
% \scalebox{1.1}{
\begin{tabular}{|c|c|c|c|c|}
\hline
% \textbf{Table}&\multicolumn{3}{|c|}{\textbf{Table Column Head}} \\
\cline{2-4} 
\textbf{Method} & \textbf{\textit{Auc}}& \textbf{\textit{F1-score}}& \textbf{\textit{Sensitivity}}& \textbf{\textit{Specificity}} \\
\hline
$Proposed$& \textbf{0.9405}& \textbf{0.8555}&\textbf{0.8655} &  0.8457\\
\hline
$Fundus_{vcdr}$& 0.9318& 0.8425& 0.7899& \textbf{0.9026}  \\
\hline
$Fundus$& 0.9219& 0.816& 0.7647& 0.8746 \\
\hline
$Mask_{vcdr}$ & 0.9042& 0.8185& 0.7647&  0.8804\\
\hline
$Mask$& 0.8916& 0.8335& 0.8067&  0.8621\\
\hline
$VCDR$&0.8949 &0.8222 & 0.7983&  0.8476\\
\hline
% \multicolumn{4}{l}{We optimize the sensitivity, specificity, and F1-score on the validation set.}
\end{tabular}
\label{tab_SY}
\end{center}
\end{table}

\begin{figure}[]
\centering
\includegraphics[scale=0.4]{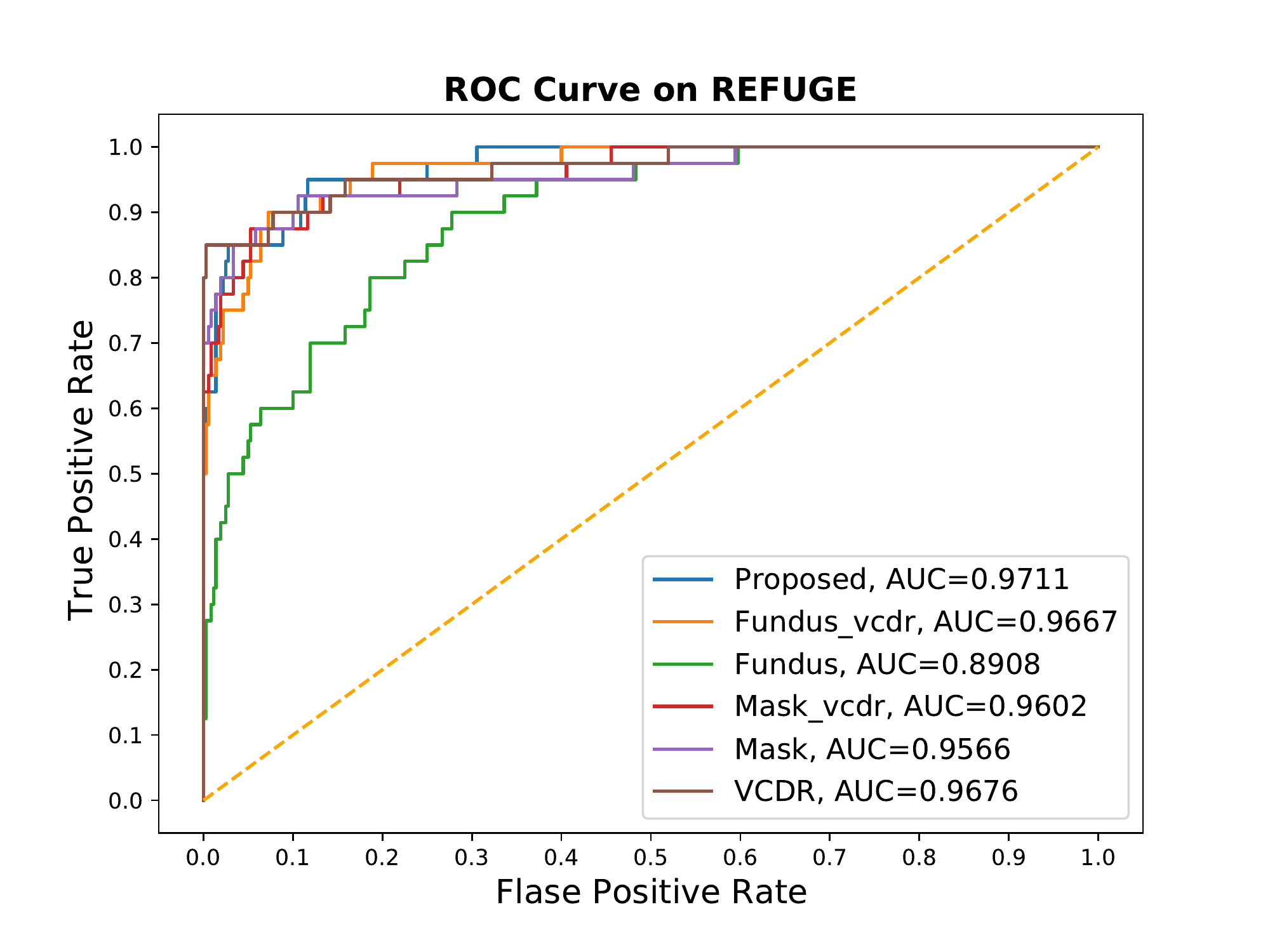}
\caption{ROC curve on REFUGE}
\label{REFUGE_roc}
\end{figure}

\begin{table}[t]
\caption{Experimental Results on REFUGE \cite{refuge}}
\begin{center}
% \scalebox{1.1}{
\begin{tabular}{|c|c|c|c|c|}
\hline
% \textbf{Table}&\multicolumn{3}{|c|}{\textbf{Table Column Head}} \\
\cline{2-4} 
\textbf{Method} & \textbf{\textit{Auc}}& \textbf{\textit{F1-score}}& \textbf{\textit{Sensitivity}}& \textbf{\textit{Specificity}} \\
\hline
$Proposed$& \textbf{0.9711}& \textbf{0.9094}& \textbf{0.95}&  0.8722\\
\hline
$Fundus_{vcdr}$ & 0.9667& 0.9032& 0.875&  \textbf{0.9333}\\
\hline
$Fundus$& 0.8908& 0.8055& 0.8&  0.8111\\
\hline
$Mask_{vcdr}$ & 0.9602& 0.8916& 0.9& 0.8833 \\
\hline
$Mask$& 0.9566& 0.9065& 0.925&  0.8889\\
\hline
$VCDR$&0.9675 &0.8993 & 0.875&  0.925\\
\hline
% \multicolumn{4}{l}{We optimize the sensitivity, specificity, and F1-score on the validation set.}
\end{tabular}
\label{tab_REFUGE}
\end{center}
\end{table}

\subsection*{Experiment On Private Dataset}
Table.~\ref{tab_SY} shows our experimental results of all the methods on the SY dataset, and the ROC curve can be observed in Fig.~\ref{SY_roc}. As we select the model with the best AUC on the validation set for each method, we also optimize the threshold values for sensitivity, specificity, and F1-score on the validation set and select the values that maximize the F1-score and assure that the specificity is greater than 0.8.

As can be observed in Table.~\ref{tab_SY}, our proposed method achieves the best Auc, F1-score, and sensitivity on SY. SY is a relatively large dataset, and our results on it shows the superiority of our method. SY can well represent the problem of long-tail distribution as well as relatively low image quality in the wild, and thus our results also reveal that the proposed method have great robustness. 

On SY, segmentation-based methods like $VCDR$, $Mask$, and $Mask_{vcdr}$ show limited performance, but combined with the rich information in fundus images, $Proposed$ and $Fundus_{vcdr}$ demonstrate better results than both segmentation-based models and the purely RGB-image-based model of $Fundus$. Besides, the $Proposed$ method also betters $Fundus_{vcdr}$ in overall performance, indicating that the two channels of new input that we construct contain information relevant to glaucoma detection. Moreover, $Fundus_{vcdr}$ also shows better performance than $Fundus$, which verifies the effectiveness of our $VCDR$ $Incorporation$ module.

\subsection*{Experiment On Public Dataset}

Table.~\ref{tab_REFUGE} shows our experimental results of all the methods on the REFUGE dataset, and the ROC curve can be observed in Fig.~\ref{REFUGE_roc}. Regarding the sensitivity, specificity, and F1-score, we adopt the same optimization method as that of SY, selecting threshold values using the same strategy for each method on the validation set of REFUGE.

Similar to SY, our proposed method also achieves the best Auc, F1-score, and sensitivity on REFUGE. We can observe that segmentation-based methods like $VCDR$, $Mask$, and $Mask_{vcdr}$ show much better performance than $Fundus$ on REFUGE, which is different from what we observe on SY. Considering that REFUGE is a much smaller dataset than SY, it is safe to speculate that segmentation-based methods are especially needed for small training sets, and more so than large ones. It can be seen that compared to $Fundus$, $Fundus_{vcdr}$ demonstrates significant improvement, which reveals that our proposed $VCDR$ $Incorporation$ module has been especially helpful on this small dataset of REFUGE, where training on a limited number of fundus images alone cannot result in good performance. Moreover, similar to what we see on SY, the $Proposed$ method betters $Fundus_{vcdr}$ in overall performance, which shows the effectiveness of our rough vessel computation and complexity reduction modules.

\section*{Conclusion}
In this paper, we propose a novel segmentation-based information extraction and amalgamation method in fundus images for the task of glaucoma detection. We manage to leverage the robustness of segmentation masks without disregarding the rich information in the original fundus images, resulting in better performances, better generalization abilities as well as image processing methods with potential for researchers in the medical field to tap into. Experimental results reveal that our method provides great improvements on both fundus-based and mask-based methods.

\section*{Acknowledgments}
This work is supported by the Fundamental Research Funds for the Central Universities, and the Research Funds of Renmin University of China (20XNA031).
%Bibliography
\bibliographystyle{unsrt}  
\bibliography{templateArxiv}

\end{document}